\documentclass[10pt]{article}
\author{Dan Elton}
\date{\today}

\usepackage{amsmath}    
\usepackage{graphicx}   
\usepackage{float}      
\usepackage{verbatim}   
\usepackage{color}      
\usepackage{hyperref}   
\usepackage{amssymb}    
\usepackage{latexsym}   
\usepackage{multirow}   
\usepackage{array}      
\usepackage{mhchem}     
\usepackage{cancel}     

\begin{document}

\title{Stretched Exponential Relaxation}


\author{Daniel C. Elton}

\date{11/2013\\ published on arXiv \today}

\maketitle
\section{The stretched exponential}
\begin{figure}[H]
  \begin{center}
    \includegraphics[width=8cm]{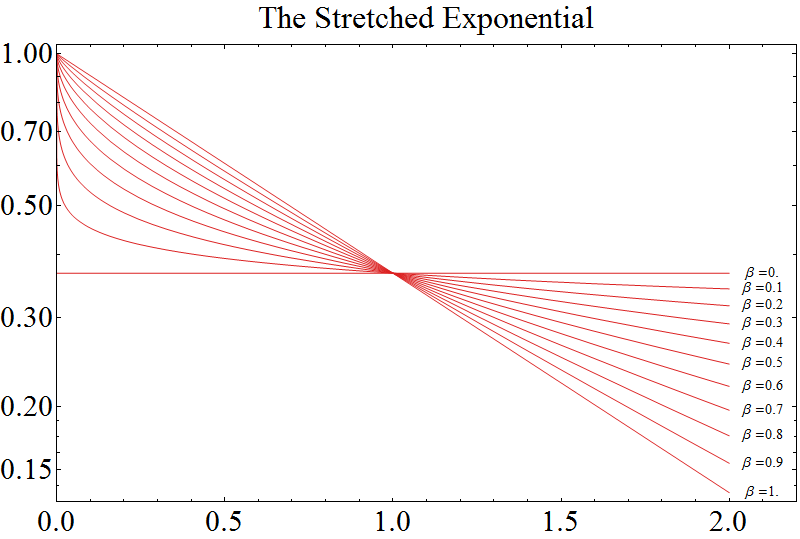}
      \caption{The stretched exponential on a log plot for various values of $\beta$}
  \end{center}
\end{figure}
The stretched exponential is:
\begin{equation}\label{StrExp}
    \phi(t) = A \exp \left[-\left(\frac{t}{\tau_{\mbox{\footnotesize{str}}}}\right)^\beta\right]
\end{equation}

Rudolf (Hermann Arndt) Kohlrausch (~1809 - 1858) first proposed the stretched exponential in 1854 to describe the relaxation of charge from a glass Leiden jar.\cite{KR:179}\footnote{{\bf Historical aside} : unfortunately a large number of works (about 2/3 of citations) do not provide the correct citation to R. Kohlrausch's work.\cite{Cardona:1521} The error seems to have stemmed from an article by Palmer, Stein, \& Abrahams in 1984 where they improperly cited an 1847 article by R. Kohlrausch.\cite{Cardona:1521} In his 1847 article R. Kohlrausch describes qualitatively the torsional relaxation of galvanometer threads but does not mention the stretched exponential expression.\cite{KR:353} The confusion on this may have stemmed from the fact that R. Kohlrausch's son, Friedrich (Wilhelm Georg) Kohlrausch (1840-1910) used the stretched exponential to describe the same mechanical relaxations, but not until a later paper in 1863.\cite{KF:1521} (Friedrich Kohlrausch also used the stretched exponential to describe mechanical stretching/creep relaxations in the natural polymer silk in 1863 and 1866.) Thus the first mention of the stretched exponential in the literature is in 1854, not 1847.\cite{KR:179} The second documented use of the stretched exponential appears to be by Werner in 1907 to describe the luminescence decay of an organic phosphor.\cite{Werner:164} In his article he does not cite Kohlrausch, suggesting his use of the stretched exponential was independent of Kohlrausch's. However, as was recently pointed out, whereas Kohlrausch's work has received thousands of citations, Werner's work has received only 8 SCI citations.\cite{Berberan:1521}. The Williams-Watts work in 1970 does not provide any citations and is done seemingly without any knowledge of earlier uses, thus lending some justification to the immense number of citations (2000+) their paper gained.  The term ``stretched exponential" does not appear in the literature until 1984.}

He came to expression \ref{StrExp} by assuming that the decay rate $ k = 1/tau$ was not constant but deceased with time as $t^{\beta -1}$.

The stretched exponential has an exceptionally wide range of application throughout science. In the materials science and engineering literature, an expression known as the ``Weibull distribution" is used , with $\beta$ taking any value. (For $\beta = 2$ the Gaussian distribution is obtained) In the social/economic sciences it is used to describe things such as urban population sizes, currency exchange rate fluctuations, and the success of Hollywood blockbusters, musicians and scientists. (Since it has a fat tail, like the power law it is useful in describing ``rich get richer" phenomena which are widespread in the social sciences). In 1993 B\"{o}mer listed stretching exponents for over 70 materials, obtained by viscoelastic, calorimetric, dielectric, optical, and other response measurements.

In mathematics, the stretched exponential is the characteristic function (Fourier transform of) a type of stable distribution called the (L\'{e}vy) ``symmetric alpha-stable distribution". (There is no general analytic expression for the Fourier transform of the stretched exponential, so this class of distributions is defined simply as the Fourier transform of the stretched exponential - this point is discussed further below).

In the field of dielectric spectroscopy it is often refereed to as the Williams-Watts (WW) or Kohlrausch-Williams-Watts (KWW) function after a paper by Williams \& Watts in 1970.\cite{Williams:80}

In all of its many applications, the stretched exponential distribution never gives the entire picture, but as a phenomenological function it can often describe data over many orders of magnitude. The fact that it contains only one extra parameter, $\beta$, makes it exceptionally easy to fit to data, but as we will see it does not have clean mathematical properties like the Gaussian or exponential function, both of which are special cases of the stretched exponential. The fact that such a simple analytic expression can fit data over many orders of magnitudes suggests, but by no means proves, that it must be of some fundamental significance.\footnote{However , there is probably not a single model explaining the presence of the stretched exponential in all these cases. This is an example of a common fallacy which physicists fall into - which is  to assume that things described by the same mathematical structure have  the same physical origin. As one example, Newtonian gravitation and  electrostatics are often said to have the same ``mathematical structure"  and thus students are left with the impression that they are somehow  the same ``in some deep way" when in fact they are vastly different both  in theory, practice, and origin (apart from the obvious difference in  the strength of the two forces, their origins are vastly different - one  comes from taking the limit of large energy density (lots of photons),  the other comes from taking the limit of small energy density (low  gravitational fields)). As a more extreme example, power laws can be  used to describe the distributions of city sizes, word frequencies and  earthquake magnitudes (to name just a few things), yet it would be silly  to assume there is common ``physics" behind them (other than the fact  that a power law describes them). In this case many disparate systems  are described by the stretched exponential, so it does not seem likely  that a common physical mechanism (or a single physical model) can be  used to explain the presence of this behavior for all of them (in the  opinion of this author). Still, just as the power law is generally  indicative of the presence of self-similarity, there are likely common  physical features associated with the stretched exponential in a very universal way.} In the field of glasses though, it can be said with good confidence that the stretched exponential has some truly fundamental significance. Even the simplest theoretical model of a glass - the 3D $\pm J$ spin glass on a cubic lattice - exhibits stretched exponential relaxation in the spin-spin autocorrelation function, as has now been conclusively shown by computer simulation.\footnote{This could be a point of contention because the spin glass model does not contain dynamics -- they must be added on top. Most of the rigorous work on spin glasses is focused on exploring their energy landscape and equilibrium properties rather than their dynamics. There are many ways of adding dynamics -- for one such dynamics -- the ``Glauber dynamics" -- theoretical arguments predict a power law decay.\cite{mezard1987}. The result quoted by Johnston is from ``molecular dynamics simulation" which give a ``overall time behavior" of $s(t) \propto t^{-x} \exp[-(t/\tau)^\beta]$, with the prefactor $t^{-x}$ being inconsequential for $t \gg 0$.\cite{Phillips:1133} The classic reference is Ogielski, 1985 -- computer simulations published before that were plagued by finite size effects due to lattice waves and did not show stretched exponential relaxation.\cite{Phillips:1133} At least $16^3 = 4096$ spins are required to get stretched exponential relaxation.\cite{Phillips:1133}} A nearly linear decrease in $\beta(T)$ as a function of temperature is observed - decreasing from $\beta = 1$ at the Curie point $\approx 4.5 T_g$ to $\beta = .35$ at the glass transition temperature $T_g$.

In the studies of relaxation of glasses, theoretical justification can be found in the so-called ``trapping model", which was first developed in the 1970s by mathematical physicists to describe the non-radiative recombination of excitons in amorphous semiconductors. The mathematical intricacies of the trapping model are quite complex, but the basic idea is rather simple to visualize. It is assumed that the material is filled with randomly distributed traps. Traps capture excitations whenever they are within a certain radius of the trap. During relaxation, excitons (or more generally, molecules) diffuse through the material and fill these traps. As the traps are filled, the remaining excitons must travel for longer to find unfilled traps, thus explaining the $k = t^{\beta - 1}$ decrease in the rate. In particular, the trapping model predicts:
\begin{equation}
    \beta = \frac{d}{d + 2}
\end{equation}
In three dimensions, the trapping model gives $\beta = 3/5 = .6$. There have been many other attempts to give the stretched exponential a theoretical justification, mostly in the field of glasses and disordered materials -- a review of these theories and the related experimental data was given by J.C. Phillips in 1996, with the focus mainly on the trapping model.\cite{Phillips:1133} According to Phillips the trapping model is the only model which describes the emergence of the stretched exponential in truly microscopic terms and does so with mathematical rigor (albeit only with great effort, requiring concepts from axiomatic set theory).\cite{Phillips:1133} Furthermore, Phillips argues that the trapping model is much more widely applicable than was previously thought.

Klafter identifies three separate models which yield a stretched exponential relaxation -- the F\"{o}rster energy transfer model, the hierarchically constrained dynamics picture and the ``defect diffusion model" -- and he shows that all three share mathematical similarities.\cite{Klafter:848}

Models yielding stretched exponential relaxation can be divided into two types - serial and parallel.\cite{Edholm:221} (Additionally, there are references in the literature suggesting that it might necessary to combine both). In a serial model, relaxation proceeds through a number of steps separated by barriers of different heights. In a parallel model, each molecule is relaxing exponentially through a single barrier, but there are different barrier heights for different molecules.

The author of this article has studied dielectic relaxation in water with molecular dynamics simulation, noting how the rotational relaxation of molecules follows a stretched exponential form.\cite{EltonThesis} Possible serial and parallel explanations for this were explored.\cite{EltonThesis}  Counter intuitvely, the overall dielectric relaxation of water is largely exponential,\cite{Elton2017:debye} but interestly the dielectric relaxation becomes stretched exponential in nano-confined water.\cite{Shekhar2014water}

\subsection{Mathematical properties of the stretched exponential}
\subsubsection{$\tau$ distribution}
\begin{figure}[H]
  \begin{center}
    \includegraphics[width=10cm]{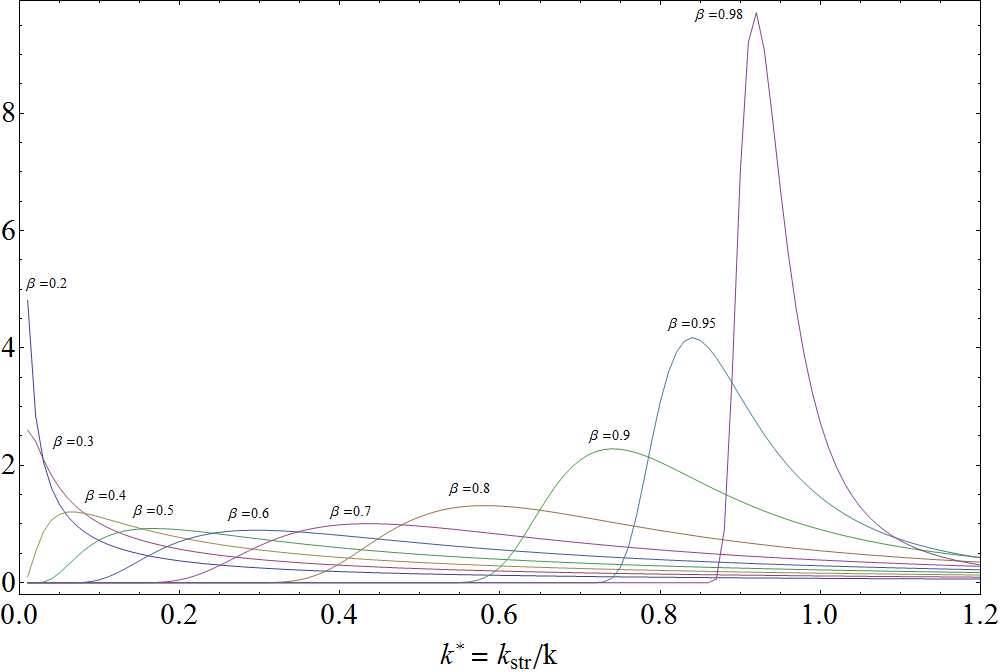}
      \caption{$G(k)$ distribution for various values of $\beta$.}
  \end{center}
    \label{kDistribution}
\end{figure}
 \begin{figure}[H]
  \begin{center}
    \includegraphics[width=10cm]{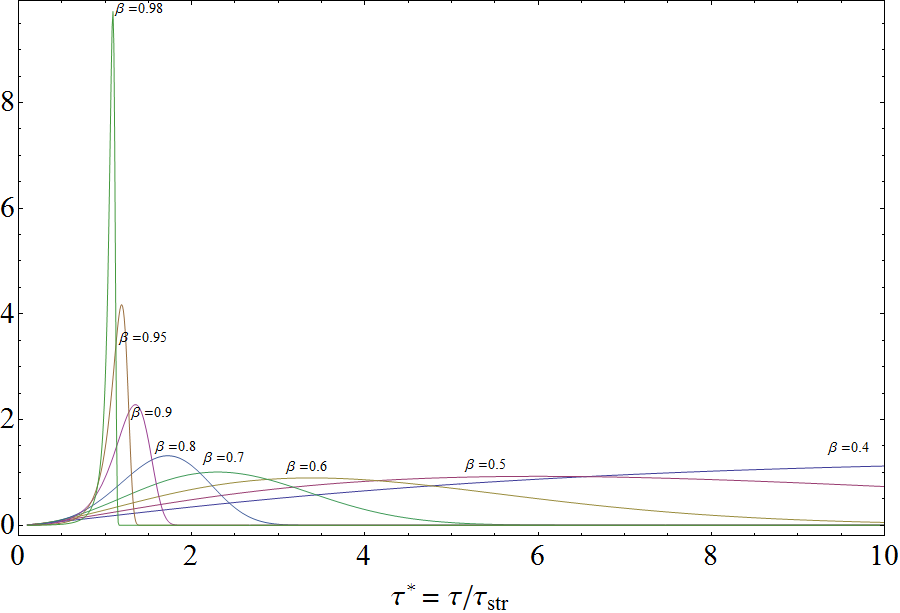}
      \caption{Same data as in fig. \ref{kDistribution} but showing $G(\tau)$ for various values of $\beta$. (It appears that the normalization was lost during the transformation)}
  \end{center}
    \label{TauDistribution}
\end{figure}

The stretched exponential can be understood as a continuous sum of exponential decays via the following transformation:
\begin{equation}\label{distributions}
    \exp \left[-\left(\frac{t}{\tau_{\mbox{\footnotesize{str}}}}\right)^\beta\right] =
        \int_0^\infty P(s,\beta) \exp ( -s \frac{t}{\tau_{\mbox{\tiny{str}}}} )ds
\end{equation}

Here $s$ is defined as the dimensionless variable $s \equiv k / k_{\mbox{\footnotesize{str}}} = \tau_{\mbox{\footnotesize{str}}} / \tau$.

Equation \ref{distributions} is an example of a Laplace transform, thus, to recover $P(s,\beta)$ requires taking the inverse Laplace transform:
\begin{equation}
    P(s,\beta) = \frac{1}{2\pi i} \int_\Gamma e^{-x^\beta} e^{sx} dx
\end{equation}
Where $\Gamma$ is a path from $-i\infty$ to $i\infty$ such that all the singularities in the integrand lie on the left.

A change of variables $ u = ix$ allows one to write this as a Fourier transform:
\begin{equation}\label{FTransformform}
    P(s,\beta) = \frac{1}{2\pi } \int_{-\infty}^{\infty} e^{-(iu)^\beta} e^{isu} du
\end{equation}

$P(s,\beta)$ is a probability distribution normalized to one. Mathematically, these distributions are asymmetric L\'{e}vy stable distributions.\cite{Jurlewicz:69} (Not to be confused with the {\it symmetric} L\'{e}vy stable distributions, of which the stretched exponential is the characteristic function). (More precisely stated, there is a one-to-one correspondance between the asymmetric L\'{e}vy stable distributions and the space of functions of the form $\exp ((kt)^\beta)$ where $\beta \in (0,2)$.) Just as the centralized limit says that idd random variables sampled from distributions with finite variance will eventually converge to a Gaussian, the generalized centralized central limit theorem says that iid variables taken from distributions with power law tails of the form $|x|^{-\beta - 1}$ ($\beta \in (0,2)$) will lead to Asymmetric L\'{e}vy stable distributions (since power law tails are not normalizable, a special normalization factor $1/N^{1/\beta}$ must be used to construct the limiting distribution).\cite{Koponen:154}  In light of this fact the ubiquitous nature of the stretched exponential is not as surprising, since power law distributions are also ubiquitous.

A few different formulae have been derived for calculating $P(s,\beta)$ for arbitrary $\beta$.  One is a summation formula:
\begin{equation}\label{slowsummationform}
    P(s,\beta) = \frac{1}{\pi} \sum_{n=1}^{\infty} \frac{ (-1)^{n+1} \Gamma (n \beta + 1) }{ n! s^{n\beta +1}} \sin ( n\pi \beta )
\end{equation}
The convergence of this series appears to be very slow -- to produce good plots $\approx$ 10,000 terms were required, thus making use of this formula very cumbersome even on a fast machine. A better equation can be obtained by looking at the real and imaginary parts of equation \ref{FTransformform}. It can be shown that the imaginary part is exactly zero (which is good because imaginary probabilities don't make much sense). The real part is:\cite{Johnston:184430}
\begin{equation} \label{fastintegralform}
    P(s,\beta) = \frac{1}{\pi} \int_0^\infty e^{-u^\beta \cos (\pi \beta / 2) } \cos [su - u^\beta \sin (\pi \beta / 2)] du
\end{equation}

A rigorous derivation of this formula is given in \cite{Liebovitch:443} using different notation. Equation \ref{fastintegralform} evaluates much faster in Mathematica (production of a nice looking plot of $P(s,\beta)$ at $\approx$ 1000 points takes about 1.5 minutes distributed over four 2.2 GHz i7 cores).

It is also possible to obtain so-called ``closed form expressions" (in terms of Airy, Bessel, Gamma, hypergeometric and complex error functions) from \label{FTransformform} for rational values of $\beta = n/m$, although these expressions rapidly increase in complexity as with larger $m$. A particularly nice closed form expression is obtained for $\beta = 1/2$:
\begin{equation}
    P(s,\frac{1}{2}) = \frac{1}{\sqrt{4\pi s^3}} \exp \left(- \frac{1}{4s}\right)
\end{equation}

The case $\beta  = 1/2$ was studied exclusively by Williams \& Watts in their seminal paper.

\subsubsection{Behavior near zero}
Near zero the stretched exponential exhibits a singularity in its derivative. In fact, this singularity appears at all higher orders of derivatives as well.
When $t \ll \tau_{\mbox{\footnotesize{str}}}$ one finds
\begin{equation}\label{nearzero}s
    \phi (t) \approx A - A\beta \left(\frac{t}{\tau_{\mbox{\footnotesize{str}}}}\right)^\beta
\end{equation}

%

\subsection{Properties of $G(k) / G(\tau)$}
The properties and limiting behavior of $P(s,\beta)$ are studied in detail by Johnson in his work to better understand how to interpret both $\beta$ and $k_{\mbox{\footnotesize{str}}}$.\cite{Johnston:184430} We will not go through his detailed analysis here, but will note some of the key results because they are very illuminating.
\begin{itemize}
    \item The max in $G(\tau)$ increases with increasing $\beta$ in a way which can be fit by an exponential function
    \item $G(k)$ contains a low-$k$ (high $\tau$) exponential cutoff that decreases with decreasing $\beta$.
    \item $G(k)$ contains a power law tail at high k described by :
    \begin{equation}
        \begin{aligned}
            P(k \rightarrow \infty) & = \frac{c}{s^{1+\beta}} \\
                c &= \frac{\Gamma(\beta + 1)}{\pi} \sin(\pi\beta)
            \end{aligned}
        \end{equation}
    \item The first moment (average) of $G(k)$ is infinite
    \item All higher moments of $G(k)$ are infinite
    \item The moments of $G(\tau)$ are finite and are given by:
        \begin{equation}
                \langle \tau^m \rangle = \frac{\Gamma(m/\beta)}{\beta \Gamma(m)}
        \end{equation}
    \item $k_{\mbox{\footnotesize{str}}}$ in the stretched exponential function is not the average rate, nor is it the inverse of the average $\tau$. It is approximately the median $k$, but only for $ .5 \beta 1$.
    \item The $\beta$ factor cannot in any way be used to describe adequately the ``width" of the distribution $G(k)$. However it is related monotonically to the low $k$ (large $\tau$) cutoff. From the average of $G(\tau)$ another interpretation is that $1/\beta$ is a measure of the average relaxation time relative to $\tau_{\mbox{\footnotesize{str}}}$.
\end{itemize}

\subsection{Barrier height distributions}

For a system where the higher energy state is separated from the lower energy state by a barrier of height $E_a$, the relaxation time (assuming Arrhenius behavior) is related to $E_a$ via:
\begin{equation}
    \begin{aligned}
        k &= k_0 e^{-\frac{E}{kT}} = \tau_0 e^{-\epsilon} \\
        E &= -kT \ln (k/k_0) = -k T \ln ( \tau_0 / \tau )
    \end{aligned}
\end{equation}
This allow one to transform $P(s,\beta)$ into $P(E_a,\beta)$, which is a distribution of barrier heights.  A way of calculating these barrier height distributions and approximating them by analytic functions was discussed by Edholm \& Blomberg in 2000.\cite{Edholm:221} These distributions look like skewed Gaussians, with skewness towards the smaller energies. In the limit of $\beta \rightarrow 0$ the distribution becomes Gaussian, but the approach is not a smooth approach - even for values of $\beta$ close to zero the distribution remains very skewed. Edholm \& Blomberg note that the approximate width of the barrier height distribution is well described by
\begin{equation}
    \sigma = 1/\beta
\end{equation}
for $ 0 < \beta < .85$. They also note that for $0 < \beta < .85$ the distribution can be somewhat approximated by a Gaussian and analyze how good this approximation is. One one might therefore wonder what happens if we just assume a Gaussian barrier height distribution. Unfortunately this leads to taking the inverse Laplace transform of an expression of the form $\exp ( \exp (x^2) )$, for which no closed form expression is known.

\bibliographystyle{plain}
\bibliography{str_exp.bib}

\begin{thebibliography}{10}

\bibitem{Berberan:1521}
M.~Berberan-Santos, E.N. Bodunov, and B.~Valeur.
\newblock History of the kohlrausch (stretched exponential) function:
  Pioneering work in luminescence.
\newblock {\em Annalen der Physik}, 17(7):460--461, 2008.

\bibitem{Cardona:1521}
M.~Cardona, R.V. Chamberlin, and W.~Marx.
\newblock The history of the stretched exponential function.
\newblock {\em Annalen der Physik}, 16(12):842--845, 2007.

\bibitem{Edholm:221}
Olle Edholm and Clas Blomberg.
\newblock Stretched exponentials and barrier distributions.
\newblock {\em Chemical Physics}, 252(1–2):221 -- 225, 2000.

\bibitem{EltonThesis}
Daniel~C. Elton.
\newblock {\em PhD Thesis : ``Understanding the Dielectric Properties of
  Water''}.
\newblock PhD thesis, Stony Brook University, 2016.

\bibitem{Elton2017:debye}
Daniel~C. Elton.
\newblock The origin of the debye relaxation in liquid water and fitting the
  high frequency excess response.
\newblock {\em Phys. Chem. Chem. Phys.}, 19:18739--18749, 2017.

\bibitem{Johnston:184430}
D.~C. Johnston.
\newblock Stretched exponential relaxation arising from a continuous sum of
  exponential decays.
\newblock {\em Phys. Rev. B}, 74:184430, Nov 2006.

\bibitem{Jurlewicz:69}
A.~Jurlewicz and K.~Weron.
\newblock A relationship between asymmetric lévy-stable distributions and the
  dielectric susceptibility.
\newblock {\em Journal of Statistical Physics}, 73:69--81, 1993.

\bibitem{Klafter:848}
Joseph Klafter and Michael~F. Shlesinger.
\newblock On the relationship among three theories of relaxation in disordered
  systems.
\newblock {\em Proceedings of the National Academy of Sciences},
  83(4):848--851, 1986.

\bibitem{KF:1521}
F.~Kohlrausch.
\newblock Ueber die elastische nachwirkung bei der torsion.
\newblock {\em Annalen der Physik}, 195(7):337--368, 1863.

\bibitem{KR:353}
R.~Kohlrausch.
\newblock {\em Pogg. Ann. Phys. Chem.}, 72:353--393, 1847.

\bibitem{KR:179}
R.~Kohlrausch.
\newblock Theorie des elektrischen rückstandes in der leidener flasche.
\newblock {\em Annalen der Physik}, 167(2):179--214, 1854.

\bibitem{Koponen:154}
Ismo Koponen.
\newblock Random transition rate model of stretched exponential relaxation.
\newblock {\em Journal of Non-Crystalline Solids}, 189(1–2):154 -- 160, 1995.

\bibitem{Liebovitch:443}
Larry~S. Liebovitch and Tibor~I. T\'{o}th.
\newblock Distributions of activation energy barriers that produce stretched
  exponential probability distributions for the time spent in each state of the
  two state reaction a⇌b.
\newblock {\em Bulletin of Mathematical Biology}, 53:443--455, 1991.

\bibitem{mezard1987}
M.~M{\'e}zard, G.~Parisi, and M.{\'A}. Virasoro.
\newblock {\em Spin Glass Theory and Beyond}.
\newblock World Scientific Lecture Notes in Physics. World Scientific, 1987.

\bibitem{Phillips:1133}
J.~C. Phillips.
\newblock Stretched exponential relaxation in molecular and electronic glasses.
\newblock {\em Reports on Progress in Physics}, 59(9):1133, 1996.

\bibitem{Shekhar2014water}
Adarsh Shekhar, Rajiv~K. Kalia, Aiichiro Nakano, Priya Vashishta, Camilla~K.
  Alm, and Anders Malthe-Sørenssen.
\newblock Universal stretched exponential relaxation in nanoconfined water.
\newblock {\em Applied Physics Letters}, 105(16):161907, 2014.

\bibitem{Werner:164}
A.~Werner.
\newblock Quantitatie messungen der an- und abklingung getrennter
  phosphorescenzbanden.
\newblock {\em Annalen der Physik}, 24:64, 1907.

\bibitem{Williams:80}
Graham Williams and David~C. Watts.
\newblock Non-symmetrical dielectric relaxation behaviour arising from a simple
  empirical decay function.
\newblock {\em Trans. Faraday Soc.}, 66:80--85, 1970.

\end{thebibliography}

\end{document}